\documentclass[trackchanges]{aastex701}

\begin{document}

\title{Magnetic Field Stratification in Active-Region Plage from Ca~{\sc{ii}} K and Ca~{\sc{ii}} $\mathbf{8542}$ \AA\ Spectropolarimetry}

\author[orcid=0009-0006-9636-5021, gname='Iñigo', sname='Juanikorena Berasategi']{Iñigo Juanikorena Berasategi}
\affiliation{Instituto de Astrofísica de Canarias, E-38205 La Laguna, Tenerife, Spain}
\affiliation{Departamento de Astrofísica, Universidad de La Laguna, E-38206 La Laguna, Tenerife, Spain}
\email{ijuanikorenab@gmail.com}

\author[orcid=0000-0001-9095-9685, gname='Ernest', sname='Alsina Ballester']{Ernest Alsina Ballester}
\affiliation{Instituto de Astrofísica de Canarias, E-38205 La Laguna, Tenerife, Spain}
\affiliation{Departamento de Astrofísica, Universidad de La Laguna, E-38206 La Laguna, Tenerife, Spain}
\email{ernest.alsina@iac.es}

\author[orcid=0000-0001-5131-4139, gname='Javier', sname='Trujillo Bueno']{Javier Trujillo Bueno}
\affiliation{Instituto de Astrofísica de Canarias, E-38205 La Laguna, Tenerife, Spain}
\affiliation{Departamento de Astrofísica, Universidad de La Laguna, E-38206 La Laguna, Tenerife, Spain}
\affiliation{Consejo Superior de Investigaciones Cient\'ificas, Spain}\email{jtb@iac.es}

\author[orcid=0000-0003-1409-1145,sname='Iglesias']{Francisco~A.~Iglesias} \affiliation{Max-Planck-Institut für Sonnensystemforschung, Justus-von-Liebig-Weg 3, 37077 Göttingen, Germany}\affiliation{Grupo de Estudios en Heliofísica de Mendoza, CONICET, Universidad de Mendoza, Boulogne sur Mer 683, 5500 Mendoza, Argentina}\email{iglesias@mps.mpg.de}

\author[orcid=0000-0001-7452-0656,sname='Kawabata']{Yusuke~Kawabata} \affiliation{National Astronomical Observatory of Japan, 2-21-1 Osawa, Mitaka, Tokyo 181-8588, Japan}\email{kawabata.yusuke@nao.ac.jp}

\author[orcid=0000-0003-4319-2009,sname='Castellanos~Durán']{Juan~Sebastián~Castellanos~Durán} \affiliation{Max-Planck-Institut für Sonnensystemforschung, Justus-von-Liebig-Weg 3, 37077 Göttingen, Germany}\email{castellanos@mps.mpg.de}

\author[orcid=0000-0001-5616-2808,sname='Kubo']{Masahito~Kubo} \affiliation{National Astronomical Observatory of Japan, 2-21-1 Osawa, Mitaka, Tokyo 181-8588, Japan}\affiliation{Department of Astronomical Science, The Graduate University for Advanced Studies (SOKENDAI), 2-21-1 Osawa, Mitaka, Tokyo 181-8588, Japan}\email{masahito.kubo@nao.ac.jp}

\author[orcid=0009-0009-4425-599X,sname='Feller']{Alex~Feller} \affiliation{Max-Planck-Institut für Sonnensystemforschung, Justus-von-Liebig-Weg 3, 37077 Göttingen, Germany}\email{feller@mps.mpg.de}

\author[orcid=0000-0002-5054-8782,sname='Katsukawa']{Yukio~Katsukawa} \affiliation{National Astronomical Observatory of Japan, 2-21-1 Osawa, Mitaka, Tokyo 181-8588, Japan}\affiliation{Department of Astronomy, The University of Tokyo, 7-3-1, Hongo, Bunkyo-ku, Tokyo 113-0033, Japan}\affiliation{Department of Astronomical Science, The Graduate University for Advanced Studies (SOKENDAI), 2-21-1 Osawa, Mitaka, Tokyo 181-8588, Japan}\email{yukio.katsukawa@nao.ac.jp}

\author[orcid=0000-0002-3418-8449,sname='Solanki']{Sami~K.~Solanki} \affiliation{Max-Planck-Institut für Sonnensystemforschung, Justus-von-Liebig-Weg 3, 37077 Göttingen, Germany}\email{solanki@mps.mpg.de}
\author[orcid=0000-0002-3387-026X,sname='del~Toro~Iniesta']{Jose~Carlos~del~Toro~Iniesta} \affiliation{Instituto de Astrofísica de Andalucía, CSIC, Glorieta de la Astronomía s/n, 18008 Granada, Spain}\affiliation{Spanish Space Solar Physics Consortium}\email{jti@iaa.es}
\author[orcid=0000-0003-1459-7074,sname='Lagg']{Andreas~Lagg} \affiliation{Max-Planck-Institut für Sonnensystemforschung, Justus-von-Liebig-Weg 3, 37077 Göttingen, Germany}\email{lagg@mps.mpg.de}
\author[orcid=0000-0002-9972-9840,sname='Gandorfer']{Achim~Gandorfer} \affiliation{Max-Planck-Institut für Sonnensystemforschung, Justus-von-Liebig-Weg 3, 37077 Göttingen, Germany}\email{gandorfer@mps.mpg.de}
\author[orcid=0000-0002-0787-8954,sname='Bernasconi']{Pietro~Bernasconi} \affiliation{Johns Hopkins University Applied Physics Laboratory, 11100 Johns Hopkins Road, Laurel, Maryland, USA}\email{pietro.bernasconi@jhuapl.edu}
\author[sname='Berkefeld']{Thomas~Berkefeld} \affiliation{Institut für Sonnenphysik (KIS), Georges-Köhler-Allee 401a, 79110 Freiburg, Germany}\email{thomas.berkefeld@leibniz-kis.de}
\author[orcid=0000-0001-6317-4380,sname='Riethmüller']{Tino~L.~Riethmüller} \affiliation{Max-Planck-Institut für Sonnensystemforschung, Justus-von-Liebig-Weg 3, 37077 Göttingen, Germany}\email{riethmueller@mps.mpg.de}

\author[orcid=0000-0001-6793-8528,sname='Naito']{Yoshihiro~Naito} \affiliation{Department of Astronomical Science, The Graduate University for Advanced Studies (SOKENDAI), 2-21-1 Osawa, Mitaka, Tokyo 181-8588, Japan}\affiliation{National Astronomical Observatory of Japan, 2-21-1 Osawa, Mitaka, Tokyo 181-8588, Japan}\email{yoshihiro.naito@grad.nao.ac.jp}

\author[orcid=0000-0001-9228-3412,sname='Álvarez-Herrero']{Alberto~Álvarez-Herrero} \affiliation{Instituto Nacional de T\'ecnica Aeroespacial (INTA), Ctra. de Ajalvir, km. 4, E-28850 Torrejón de Ardoz, Spain}\affiliation{Spanish Space Solar Physics Consortium}\email{alvareza@inta.es}
\author[orcid=0000-0003-3490-6532,sname='Smitha']{H.~N.~Smitha} \affiliation{Max-Planck-Institut für Sonnensystemforschung, Justus-von-Liebig-Weg 3, 37077 Göttingen, Germany}\email{narayanamurthy@mps.mpg.de}
\author[orcid=0000-0001-8829-1938,sname='Orozco~Suárez']{David~Orozco~Suárez} \affiliation{Instituto de Astrofísica de Andalucía, CSIC, Glorieta de la Astronomía s/n, 18008 Granada, Spain}\affiliation{Spanish Space Solar Physics Consortium}\email{orozco@iaa.es}
\author[sname='Grauf']{Bianca~Grauf} \affiliation{Max-Planck-Institut für Sonnensystemforschung, Justus-von-Liebig-Weg 3, 37077 Göttingen, Germany}\email{grauf@mps.mpg.de}
\author[sname='Carpenter']{Michael~Carpenter} \affiliation{Johns Hopkins University Applied Physics Laboratory, 11100 Johns Hopkins Road, Laurel, Maryland, USA}\email{michael.carpenter@jhuapl.edu}
\author[sname='Bell']{Alexander~Bell} \affiliation{Institut für Sonnenphysik (KIS), Georges-Köhler-Allee 401a, 79110 Freiburg, Germany}\email{albe@leibniz-kis.de}
\author[orcid=0000-0001-7764-6895,sname='Martínez~Pillet']{Valentín~Martínez~Pillet} \affiliation{Instituto de Astrofísica de Canarias, Vía Láctea, s/n, E-38205 La Laguna, Spain}\affiliation{Spanish Space Solar Physics Consortium}\email{vmpillet@iac.es}

\author[orcid=0000-0002-7318-3536,sname='Bailén']{Francisco~Javier~Bailén} \affiliation{Instituto de Astrofísica de Andalucía, CSIC, Glorieta de la Astronomía s/n, 18008 Granada, Spain}\affiliation{Spanish Space Solar Physics Consortium}\email{fbailen@iaa.es}
\author[orcid=0000-0002-2055-441X,sname='Blanco~Rodríguez']{Julian~Blanco~Rodríguez} \affiliation{Universitat de Valencia Catedrático José Beltrán 2, E-46980 Paterna-Valencia, Spain}\affiliation{Spanish Space Solar Physics Consortium}\email{julian.blanco@uv.es}
\author[orcid=0009-0002-6808-5154,sname='Harnes']{Edvarda~Harnes} \affiliation{Max-Planck-Institut für Sonnensystemforschung, Justus-von-Liebig-Weg 3, 37077 Göttingen, Germany}\email{harnes@mps.mpg.de}
\author[orcid=0000-0001-6029-7529,sname='Hölken']{Johannes~Hölken} \affiliation{Max-Planck-Institut für Sonnensystemforschung, Justus-von-Liebig-Weg 3, 37077 Göttingen, Germany}\email{hoelken@mps.mpg.de}
\author[orcid=0000-0002-4669-5376,sname='Ishikawa']{Ryohtaroh~T.~Ishikawa} \affiliation{National Institute for Fusion Science, 322-6 Oroshi-cho, Toki City 509-5292, Japan}\email{ishikawa.ryohtaro@nifs.ac.jp}
\author[orcid=0000-0002-1043-9944,sname='Matsumoto']{Takuma~Matsumoto} \affiliation{Centre for Integrated Data Science, Institute for Space-Earth Environmental Research, Nagoya University, Furocho, Chikusa-ku, Nagoya, Aichi 464-8601, Japan}\email{takuma.matsumoto@gmail.com}
\author[orcid=0000-0002-7044-6281,sname='Oba']{Takayoshi~Oba} \affiliation{Advanced Research Center for Space Science and Technology, Institute of Science and Engineering, Kanazawa University, Kakuma-machi, Kanazawa, Ishikawa 920-1192, Japan}\affiliation{Max-Planck-Institut für Sonnensystemforschung, Justus-von-Liebig-Weg 3, 37077 Göttingen, Germany}\email{oba@mps.mpg.de}
\author[orcid=0000-0003-0175-6232,sname='Siu-Tapia']{Azaymi~L.~Siu-Tapia} \affiliation{Instituto de Astrofísica de Andalucía, CSIC, Glorieta de la Astronomía s/n, 18008 Granada, Spain}\affiliation{Spanish Space Solar Physics Consortium}\email{siu@iaa.es}
\author[orcid=0000-0003-1483-4535,sname='Strecker']{Hanna~Strecker} \affiliation{Instituto de Astrofísica de Andalucía, CSIC, Glorieta de la Astronomía s/n, 18008 Granada, Spain}\affiliation{Spanish Space Solar Physics Consortium}\email{streckerh@iaa.es}
\author[orcid=0000-0003-1971-5551,sname='Vukadinović']{Dušan~Vukadinović} \affiliation{Institut für Physik, Universität Graz, Universitätsplatz 5, 8010 Graz, Austria}\affiliation{Max-Planck-Institut für Sonnensystemforschung, Justus-von-Liebig-Weg 3, 37077 Göttingen, Germany}\email{dusan.vukadinovic@uni-graz.at}



\begin{abstract}
We investigate the height variation of the line-of-sight (LOS) magnetic field in solar active-region (AR) plage from the upper photosphere to the upper chromosphere using co-spatial ultraviolet (UV) and infrared (IR) spectropolarimetric observations from the Sunrise III stratospheric balloon flight. The Ca~{\sc{ii}} K and Ca~{\sc{ii}} $8542$ \AA\ lines provide complementary chromospheric diagnostics, while nearby Fe~{\sc{i}} lines sample photospheric layers. The LOS magnetic field is inferred from the intensity and circular polarization profiles, applying the weak field approximation for the Ca~{\sc{ii}} lines and the center-of-gravity method for the Fe~{\sc{i}} lines. The photospheric Fe~{\sc{i}} lines reveal strong, finely structured magnetic fields of the order of a kG. In contrast, the chromospheric Ca~{\sc{ii}} diagnostics yield systematically weaker fields, typically in the $\sim 100$--$400$~G range, with a more diffuse spatial distribution. Magnetic field maps inferred from the Ca~{\sc{ii}} K line, formed higher in the chromosphere, are smoother and more extended than those derived from the Ca~{\sc{ii}} $8542$ \AA\ line, which samples lower heights. We find that the magnetized area increases by a factor of $\sim 2$ from the photosphere to the chromosphere. These results provide direct quantitative evidence that magnetic fields in AR plage weaken and expand with height, evolving from compact kG photospheric concentrations into weaker and more spatially extended structures in the upper chromosphere.
\end{abstract}



\section{Introduction} \label{Section_intro}

The magnetic field in the solar chromosphere plays a key role in several outstanding problems in solar physics, including the origin of the solar wind, the heating of the overlying corona, and the onset of eruptive phenomena. However, obtaining quantitative information on chromospheric magnetic fields remains challenging. This difficulty arises from the limited number of suitable spectral lines, the complexity of their modeling, and the relatively weak magnetic field strengths compared with the photosphere, which result in low-amplitude Zeeman polarization signals \citep{Review_TB_dPA}. Understanding the magnetic field stratification in active regions (ARs) is particularly important, as these are the sites where most energetic solar phenomena originate.\\

To probe the chromospheric magnetic field at different heights in an AR, we analyze spectropolarimetric signals from the Ca~{\sc{ii}} K and Ca~{\sc{ii}} $8542$ \AA\ lines, which provide complementary diagnostics of the upper and middle chromosphere, respectively. Early ground-based observations of the Ca~{\sc{ii}} K line at $3933$\AA\ already demonstrated its diagnostic potential for the upper chromosphere, despite limitations imposed by atmospheric seeing and instrumental constraints. In particular, \citet{K_line_1990} reported the first detection of circular polarization in this line in a sunspot penumbra. Recently, theoretical studies have investigated the magnetic sensitivity of this line and the physical processes that shape its polarization signals, further highlighting its diagnostic value for probing magnetic fields of the upper chromosphere (see \citealt{IJB_2025, Kriginsky_2026}).\\

The Ca~{\sc{ii}} $8542$ \AA\ line is likewise well suited for magnetic diagnostics in the middle chromosphere \citep[e.g.,][]{Hector-Javier-Basilio-SCIENCE,Manso-TrujilloBueno-2010,Manso-TrujilloBueno-2011,QuinteroNoda+16}. For magnetic field strengths typical of ARs, on the order of a few hundred gauss, both lines exhibit circular polarization signals at the level of a few percent of the intensity \citep{IJB_2025}. This enables an estimate of the line-of-sight (LOS) magnetic field, which in this work is inferred using the weak-field approximation (WFA; see, e.g., \citealt{Landi_2004}) in the Ca~{\sc{ii}} lines. The WFA has been extensively tested for the Ca~{\sc{ii}} $8542$ \AA\ line (e.g., \citealt{Centeno_WFA, Morosin_2020, Vissers_2022}) and was recently validated for the Ca~{\sc{ii}} K line \citep{Kriginsky_2026}. To characterize the underlying photospheric magnetic field, we additionally use the nearby Fe~{\sc{i}} $3937$ \AA\ and $8515$ \AA\ lines, which sample deeper atmospheric layers. Because stronger magnetic fields are expected at such depths, we instead apply the center-of-gravity method (COG; see, e.g., \citealt{Landi_2004}, \citealt{COG_Rees_Semel}) to those Fe~{\sc{i}} lines.\\

In recent years, spectropolarimetric studies have emerged as a powerful tool for characterizing the magnetic field across different heights in the solar atmosphere. Ground-based observations of the Ca~{\sc{ii}} $8542$ \AA\ line have been widely used to investigate the magnetic field in the middle chromosphere (e.g., \citealt{Morosin_2020, Pietrow_2020}). An important step forward was the development of multi-line spectropolarimetric diagnostics in the ultraviolet through the CLASP2 and CLASP2.1 missions, which combined the Mg~{\sc{ii}} h\&k resonance lines with other ultraviolet (UV) diagnostics. By applying the WFA to the observed Stokes $I$ and $V$ profiles, these observations enabled the chromospheric magnetic field stratification in active-region plage to be inferred \citep{CLASP_2021_Ryoko, CLASP_2025_Ryoko, Afonso_CLASP_2025}. These UV diagnostics are sensitive to the magnetic field from the photosphere through the chromosphere and into its highest layers. Owing to the higher formation heights of the Mg~{\sc{ii}} h\&k line cores, they sample atmospheric layers significantly closer to the base of the corona than those accessible with the Ca~{\sc{ii}} H and K lines, thereby extending the height range over which the magnetic field can be constrained \citep[e.g., Figure 1 in][]{STiC_de_la_Cruz}. Motivated by these advances, we adopt a complementary multi-line approach based on the Fe~{\sc{i}} and Ca~{\sc{ii}} lines to provide robust constraints on the variation of $B_{\mathrm{LOS}}$ from the photosphere to the upper chromosphere.\\

The present investigation is based on data obtained during the Sunrise III balloon flight in July 2024 \citep{Sunrise_III_overview, Sunrise_III_Solanki}, which builds on the original Sunrise observatory \citep{Sunrise} that flew successfully in 2009 and 2013 \citep{Sunrise_I, Sunrise_II}. The Sunrise UV Spectropolarimeter and Imager (SUSI; \citealt{Sunrise_III_SUSI, Sunrise_III_SUSI_spectropol}) and the Sunrise Chromospheric Infrared spectro-Polarimeter (SCIP; \citealt{Sunrise_III_SCIP}) acquired spectropolarimetric data of unprecedented quality. SUSI and SCIP provide the first spatially resolved spectropolarimetric observations of the Ca~{\sc{ii}} K and $8542$ \AA\ lines obtained from a stratospheric platform, respectively. This combination of spectral diagnostics and observing conditions offers a unique opportunity to investigate the vertical structure of magnetic fields in AR plage with unprecedented accuracy.\\

\section{Observational data and methods} \label{Section_data}

\begin{figure*}[htbp!]
    \centering
    \includegraphics[width=1.\textwidth]{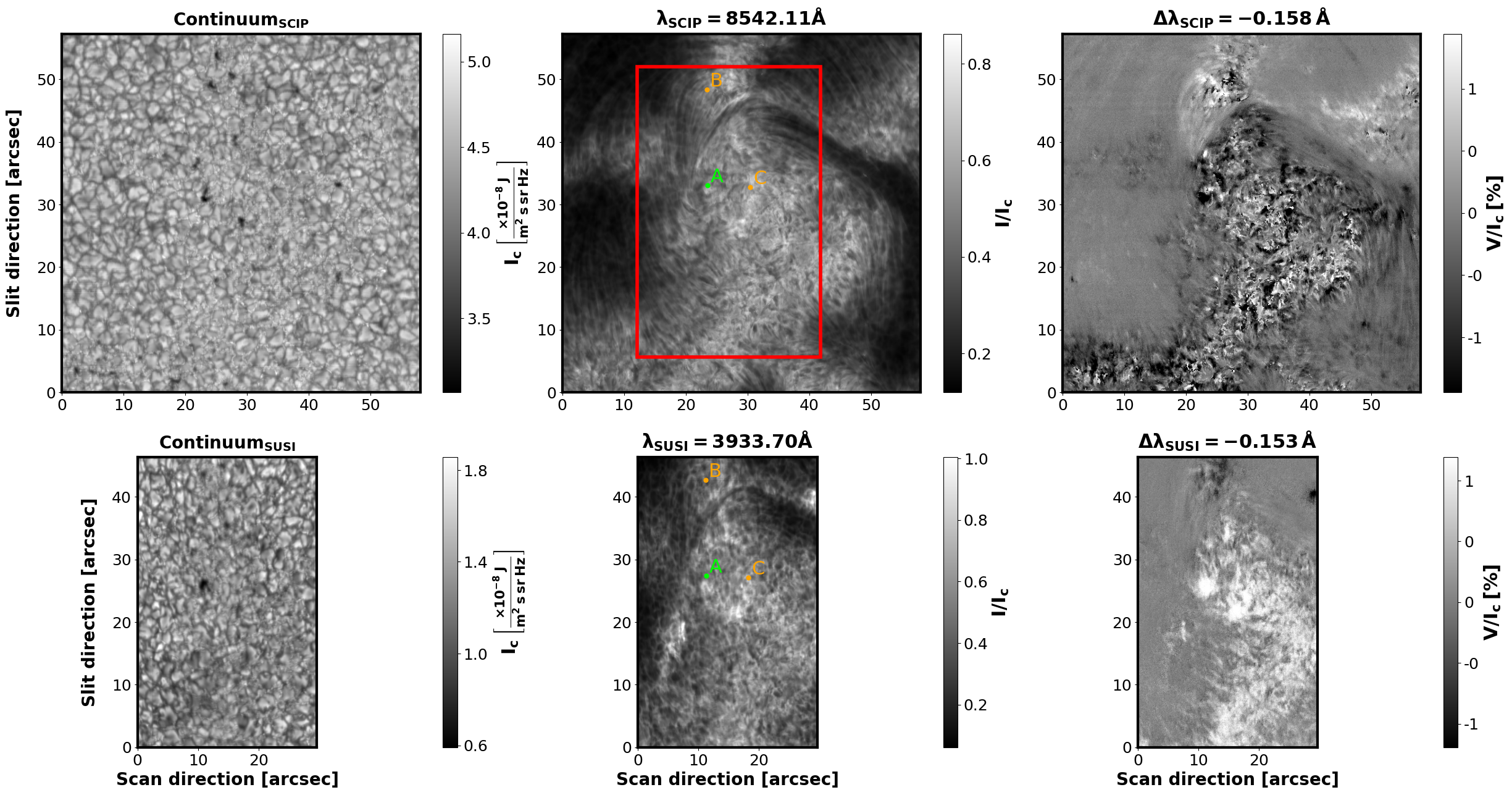}
    \caption{Images of the intensity and circular polarization at selected wavelengths of the analyzed datasets. Upper panels correspond to the SCIP observation, and lower ones to SUSI's. Left panels display intensity maps close to the continuum (around $3946$ \AA\ in SUSI and $8505$ \AA\ in SCIP), with the intensity given in SI units. Central panels show the core of the Ca~{\sc{ii}} lines of interest (K in SUSI and $8542$ \AA\ in SCIP), normalized by the local continuum. The right panels display the circular polarization in the blue wing. The wavelength considered in each panel is indicated above it, with $\Delta \lambda$ being the separation from the line core. The red rectangle indicates the position of the SUSI observation within the FOV of SCIP.}
    \label{SCIP_32_PLAG_SUSI_32_PLAG_1_stokes_maps}
\end{figure*}

We analyze two raster scans of the same solar region within AR 13745, obtained on 15 July 2024, one acquired with SUSI and the other with SCIP. The data has been given the Sunrise ID 33\_ARFL (see Table 1 of \citealt{Sunrise_III_Solanki}). The observed field of view contains an AR plage with an overlying filament that appears dark in intensity, in the core of both the Ca~{\sc{ii}} lines, K (SUSI) and Ca~{\sc{ii}} $8542$ \AA\ (SCIP) spectral lines. An analysis of the vector magnetic field of the filament using only SCIP data is presented in \citet{Matsumoto_letter}. Figure~\ref{SCIP_32_PLAG_SUSI_32_PLAG_1_stokes_maps} shows maps of the continuum intensity, the line-core intensity of the Ca~{\sc{ii}} lines normalized to the local continuum ($I/I_{\mathrm{c}}$), and the circular polarization in the wings ($V/I_{\mathrm{c}}$). We note that we take as the continuum intensity the value at $3946$ \AA\ for the SUSI data. This is still within the wings of the Ca II K line, but its intensity should be close to that of the continuum.\\

The SUSI scan started at 17:43 UT and lasted 56 minutes, whereas the SCIP scan began at 18:39 UT with a duration of 11 minutes. The datasets are co-spatial but not co-temporal, and some evolution of the observed structures may therefore occur between the two acquisitions.\\

Both instruments measured the full Stokes vector as a function of wavelength, although in this work we restrict the analysis to the intensity $I$ and circular polarization $V$. The linear polarization signals in these lines are very weak near the solar disk center, and their analysis would require spatial averaging, which we avoid here. The SUSI observations were obtained in the $393.3 \: \mathrm{nm}$ channel, which includes the Ca~{\sc{ii}} K line at $3933$ \AA\ [$g_{\mathrm{eff}} = 1.17$]. The spectral range covered by SCIP contains the Ca~{\sc{ii}} $8542$ \AA\ line [$g_{\mathrm{eff}} = 1.1$]. In addition to these chromospheric diagnostics, we also analyze two nearby Fe~{\sc{i}} lines, $3937$ \AA\ in SUSI [$g_{\mathrm{eff}} = 2.0$] and $8515$ \AA\ in SCIP [$g_{\mathrm{eff}} = 0.75$], which probe deeper atmospheric layers. We selected these lines because they are unblended, sufficiently strong, and exhibit clear $V$ signals.\\

The spectral sampling of SUSI is approximately four times higher than that of SCIP, with samplings of $\Delta \lambda \approx 0.0102$ \AA/pix and $\Delta \lambda \approx 0.0395$ \AA/pix, respectively. The spatial sampling of SUSI is about a factor of $\sim 3.2$ higher than that of SCIP in both the slit and scanning directions, with pixel sizes of $\approx 0.0296''$ for SUSI and $\approx 0.094''$ for SCIP. The integration times per slit position are $3.3$s for SUSI and $1$s for SCIP. The full field of view (FOV) is $46 \times 29''$ for SUSI and $58 \times 58''$ for SCIP, with the SUSI FOV fully contained within that of SCIP (see red rectangles in Figures~\ref{SCIP_32_PLAG_SUSI_32_PLAG_1_stokes_maps} and \ref{SCIP_32_PLAG_SUSI_32_PLAG_1_B_los_WFA_maps}).\\

To facilitate a direct comparison between the datasets, we co-aligned the images by maximizing the cross-correlation between the Ca~{\sc{ii}} $8542$ \AA\ intensity map (upper central panel in Figure~\ref{SCIP_32_PLAG_SUSI_32_PLAG_1_stokes_maps}) and the Ca~{\sc{ii}} K intensity map (lower central panel). For this purpose, the SUSI data were spatially rebinned to match the pixel size of SCIP.\\

The LOS magnetic field component ($B_{\mathrm{LOS}}$) was inferred from the intensity and circular polarization profiles of the selected spectral lines. At each spatial pixel, we applied the WFA to the Ca~{\sc{ii}} lines and the COG method to the Fe~{\sc{i}} ones (see Appendix~\ref{Appendix_COG}), with the different lines providing estimates of the magnetic field at different atmospheric heights (see Appendix~\ref{Appendix_formation_heights}). When applying the WFA, the value of $B_{\mathrm{LOS}}$ that best reproduces the observed relation between $V$ and the wavelength derivative of the intensity $\left(\frac{\partial I}{\partial \lambda}\right)$ was determined using a Markov chain Monte Carlo (MCMC) approach (see Appendix~\ref{Appendix_MCMC} for details). \\

We also computed the magnetized area in the $B_{\mathrm{LOS}}$ map corresponding to each spectral line. A pixel was classified as magnetized if its $B_{\mathrm{LOS}}$ value exceeded 20\% of the maximum longitudinal field strength, in a similar approach to that taken by \citet{CLASP_2025_Ryoko}. Because the maximum longitudinal field strengths differ substantially between positive and negative polarities, the maximum strength and the corresponding cutoff value were determined independently for each polarity after excluding the top 2\% of values to mitigate the effect of outliers. We restricted the analysis to the FOV common to both instruments in order to exclude quiet regions surrounding the plage and to ensure a more consistent comparison. The resulting cutoff field strengths are $B_{\mathrm{cutoff}} = (53, -193)$, $(93, -179)$, $(68, -127)$, and $(46, -94)$ G for Fe~{\sc{i}} $8515$ \AA, Fe~{\sc{i}} $3937$ \AA, Ca~{\sc{ii}} $8542$ \AA, and Ca~{\sc{ii}} K, respectively.

\section{Results and discussion} \label{Section_results}
\begin{figure*}[htbp!]
    \centering
    \includegraphics[width=0.85\textwidth]{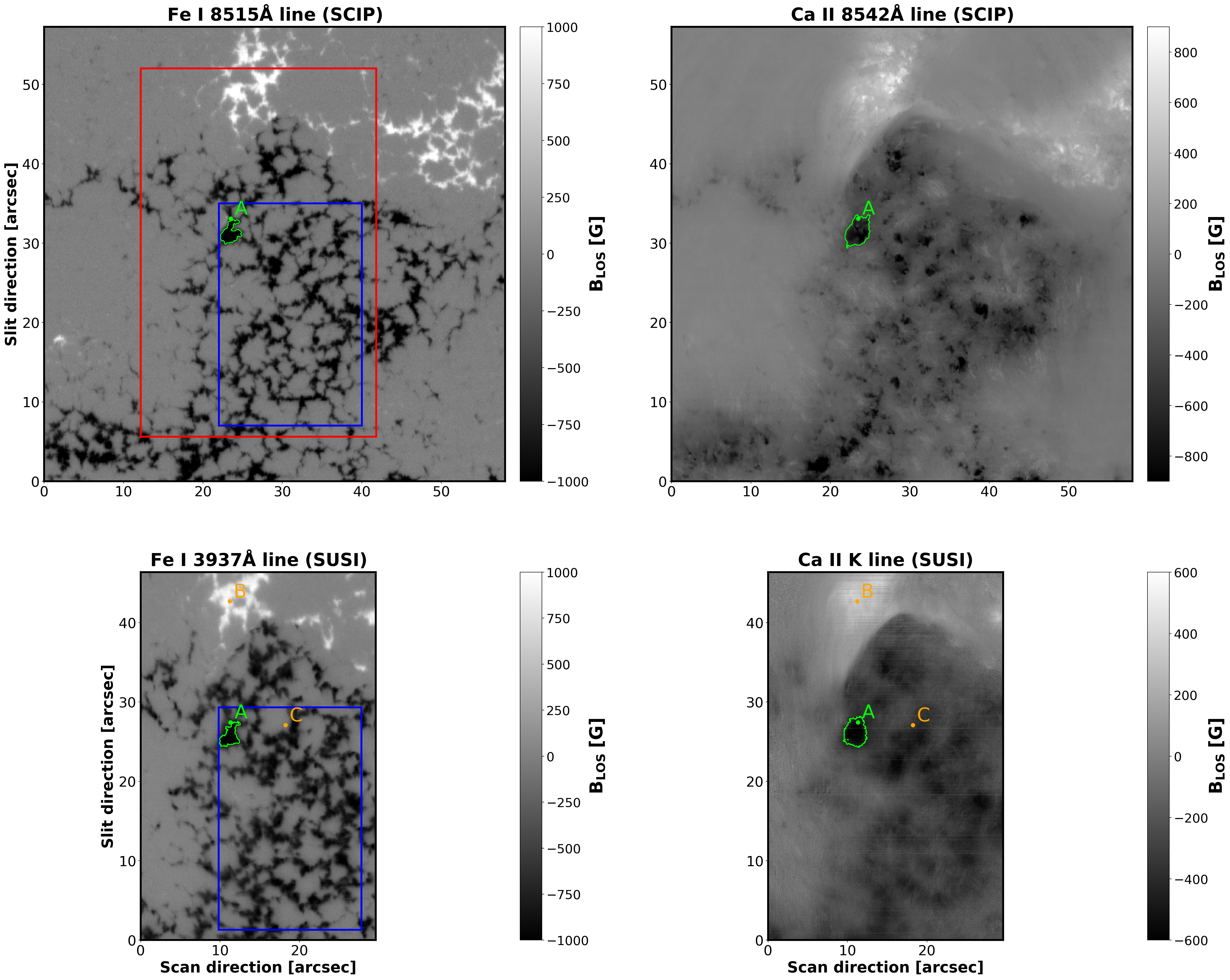}
    \caption{Magnetic field strength along the line of sight, $B_{\mathrm{LOS}}$, obtained by applying the COG method to the Fe~{\sc{i}} spectral lines (left panels), and the WFA to the Ca~{\sc{ii}} lines (right panels), in the SCIP (upper panels) and SUSI (lower panels) datasets. The red rectangle indicates the position of the SUSI observation within the FOV of SCIP. This region is used to compute the magnetized area fraction in each map. The blue rectangle indicates the predominantly negative polarity region analyzed later in this section. The green contours in the Fe~{\sc{i}} panels delimit a specific magnetic patch in the FOV where $B_{\mathrm{LOS}} > 800 \: \mathrm{G}$, and the contours in the Ca~{\sc{ii}} panels enclose the same magnetic flux as their Fe~{\sc{i}} counterparts.}
    \label{SCIP_32_PLAG_SUSI_32_PLAG_1_B_los_WFA_maps}
\end{figure*}

We inferred the longitudinal component of the magnetic field for all pixels in the FOV for the spectral lines of interest, namely Fe~{\sc{i}} $8515$ \AA\ and Ca~{\sc{ii}} $8542$ \AA\ in the SCIP data, and Fe~{\sc{i}} $3937$ \AA\ and Ca~{\sc{ii}} K in the SUSI data. The resulting $B_{\mathrm{LOS}}$ maps are shown in Figure~\ref{SCIP_32_PLAG_SUSI_32_PLAG_1_B_los_WFA_maps}.\\

The Fe~{\sc{i}} lines sample the upper photosphere at slightly different heights. In semi-empirical atmospheric models of AR plages, the $3937$ \AA\ line (lower left panel) forms around $300 \: \mathrm{km}$, higher than the $8515$\AA\ line (upper left panel), which forms at approximately $200 \: \mathrm{km}$ (see the analysis on indicative formation regions for each line and the corresponding figure in Appendix~\ref{Appendix_formation_heights}). The inferred $B_{\mathrm{LOS}}$ maps show similar large scale structures, with well-defined magnetic network features, and span the $[-1000, 1000] \: \mathrm{G}$ range. However, the largest values in this range exceed the generous upper limits for the applicability of the WFA presented in Appendix~\ref{Appendix_formation_heights} and should be treated with caution. The map derived from $8515$ \AA\ appears sharper than the one from $3937$ \AA , and their magnetized areas occupy 33\% and 49\% of the shared FOV, respectively. Such differences are consistent with the deeper formation height of $8515$ \AA . Small-scale differences between the maps likely reflect the temporal evolution between the scans, in addition to the different formation heights of the lines. In strong-field regions, $8515$ \AA\ generally yields stronger fields than $3973$ \AA\ by a few tens of gauss, consistent with the former line's deeper formation and higher magnetic field threshold for the validity of the WFA.\\

The chromospheric Ca~{\sc{ii}} $8542$ \AA\ line (upper right panel) yields weaker magnetic fields and more diffuse spatial structures than the Fe~{\sc{i}} lines, as expected from its formation in the middle chromosphere (see figure in Appendix ~\ref{Appendix_formation_heights}). Despite this smoothing, compact patches of relatively strong magnetic field remain visible within the AR plage. In the FOV shared with SUSI, 70\% of the area is classified as magnetized.\\

The Ca~{\sc{ii}} K line (lower right panel) probes higher chromospheric layers. The inferred longitudinal field strengths lie in the range $[-600, 500] \: \mathrm{G}$ and display structures broadly similar to those obtained from Ca~{\sc{ii}} $8542$ \AA\ ($[-800, 800] \: \mathrm{G}$ range), but are significantly more diffuse, with more extended and less sharply defined magnetic patches, such that 80\% of the area is classified as magnetized. According to the formation-height analysis in Appendix~\ref{Appendix_formation_heights}, both Ca~{\sc{ii}} lines sample broad height ranges, as the selected wavelength intervals include contributions from both inner and outer polarization lobes. Isolating only the inner lobes is generally not possible across the full FOV due to S/N limitations (Appendix~\ref{Appendix_wave_int_range}), so the inferred $B_{\mathrm{LOS}}$ values represent averages over these contributing layers. Nevertheless, since Ca~{\sc{ii}} K forms systematically higher than Ca~{\sc{ii}} $8542$ \AA, it consistently yields weaker and more spatially diffuse magnetic fields, despite the higher spatial resolution of SUSI.\\

Considering the FOV shared by both instruments, the magnetized area (see Section \ref{Section_data}) increases by a factor of 2.1 from the photosphere to the chromosphere for the SCIP lines, while the SUSI lines yield a slightly lower factor of 1.7. We recall that both SUSI lines form higher in the atmosphere than their SCIP counterparts, so the corresponding expansion factor is measured relative to a higher photospheric layer. These values are lower than those reported by \cite{CLASP_2025_Ryoko} for a plage region underlying a moss, where UV Mn~{\sc{i}} lines (lower chromosphere) and the outer lobes of Mg~{\sc{ii}} h\&k (middle chromosphere) yielded expansion factors of $2.7$ and $3.1$, respectively. In that study, less than 20\% of the deepest layer was classified as magnetized, compared to 33\% and 49\% in the photospheric layers analyzed here. This larger photospheric magnetization factor, together with differences in the formation heights of the spectral lines and the higher spatial resolution of the Sunrise III observations, may account for the lower expansion factors found in our analysis.\\

We additionally took a parallel approach to determine the expansion factor for a strongly magnetized patch. In the $B_{\mathrm{LOS}}$ maps obtained from Fe~{\sc{i}} lines, we selected the patch around point A shown in Figure \ref{SCIP_32_PLAG_SUSI_32_PLAG_1_B_los_WFA_maps} (green contours in left panels) by tracing a $B_{\mathrm{LOS}} > 800 \: \mathrm{G}$ contour. For the Fe~{\sc{i}} $8515$ \AA\ and $3937$ \AA\ lines, these enclose magnetic fluxes of $3.04 \times 10^{19}$ Mx and $2.02 \times 10^{19}$ Mx, respectively. We then determined the areas in the Ca~{\sc{ii}} maps that enclose the same flux by applying a region-growing algorithm. For the SCIP and SUSI instruments, such areas are larger than the corresponding patches in the Fe~{\sc{i}} maps by factors of $2.0$ and $1.7$, respectively. These expansion factors are in good agreement with those obtained through the magnetized-area analysis discussed above.\\

\begin{figure*}[htbp!]
    \centering
    \includegraphics[width=0.85\textwidth]{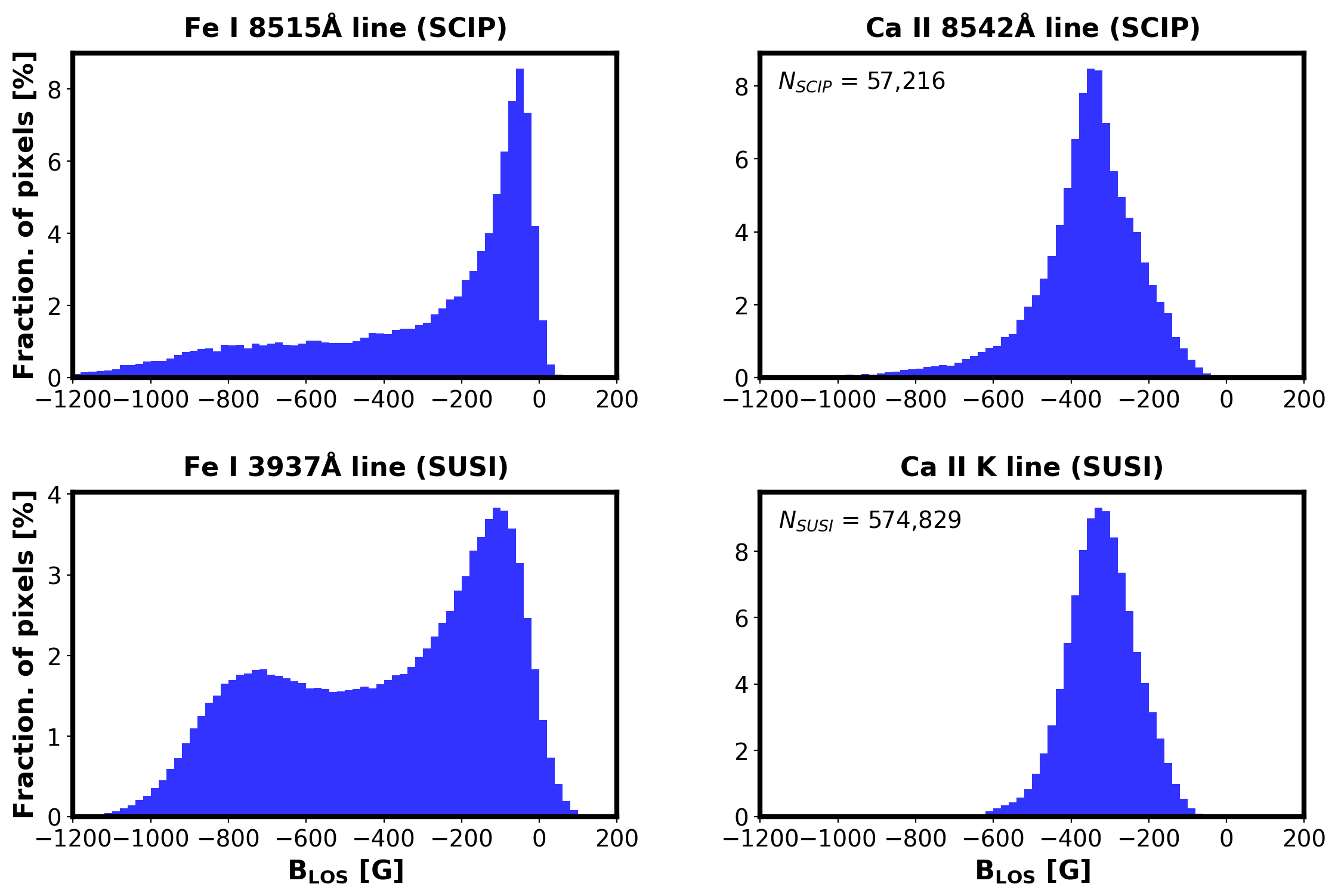}
    \caption{Distribution of the $B_{\mathrm{LOS}}$ values as inferred with each spectral line in the predominantly unipolar region enclosed by the blue rectangles in Figure \ref{SCIP_32_PLAG_SUSI_32_PLAG_1_B_los_WFA_maps}. The bin size is $20 \: \mathrm{G}$, and the number of pixels considered for each instrument is given in the right panels.}
    \label{SCIP_32_PLAG_SUSI_32_PLAG_1_B_los_histograms}
\end{figure*}

The histograms in Figure \ref{SCIP_32_PLAG_SUSI_32_PLAG_1_B_los_histograms} illustrate the magnetic field expansion in the predominantly (negative) unipolar region indicated by the blue rectangle in Figure \ref{SCIP_32_PLAG_SUSI_32_PLAG_1_B_los_WFA_maps}. The photospheric Fe~{\sc{i}} lines exhibit a broad distribution of $B_{\mathrm{LOS}}$ values, with a large number of pixels associated with weak fields in the parts of the FOV not covered by strong fields, while still retaining a substantial fraction of pixels with strong magnetic fields corresponding to the network. Among the two photospheric lines, the deeper-forming Fe~{\sc{i}} $8515$ \AA\ line displays a more extended tail toward larger field strengths.\\

The Fe~{\sc{i}} $3937$ \AA\ line observed by SUSI, despite forming about $100$~km higher than its SCIP counterpart, exhibits a two times larger fraction of pixels with opposite (positive) polarity. This is likely due to the fact that the pixel size of SUSI is approximately ten times smaller than that of SCIP, which allows it to resolve magnetic structures that remain unresolved in SCIP observations.\\

The distributions derived from the chromospheric lines are considerably narrower, consistent with a magnetic field that has expanded with height and become more spatially homogeneous. The Ca~{\sc{ii}} $8542$ \AA\ distribution exhibits a tail that extends to larger $|B_{\mathrm{LOS}}|$ values than that of the Ca~{\sc{ii}} K line. The distributions peak at $-350$ G and $-330$ G, respectively. The inference of slightly larger values from Ca~{\sc{ii}} $8542$ \AA\ than from Ca~{\sc{ii}} K is consistent with the different formation regions of the two lines, with the latter line sampling higher layers of the chromosphere.\\

\begin{figure*}[htbp!]
    \centering
    \includegraphics[width=0.8\textwidth]{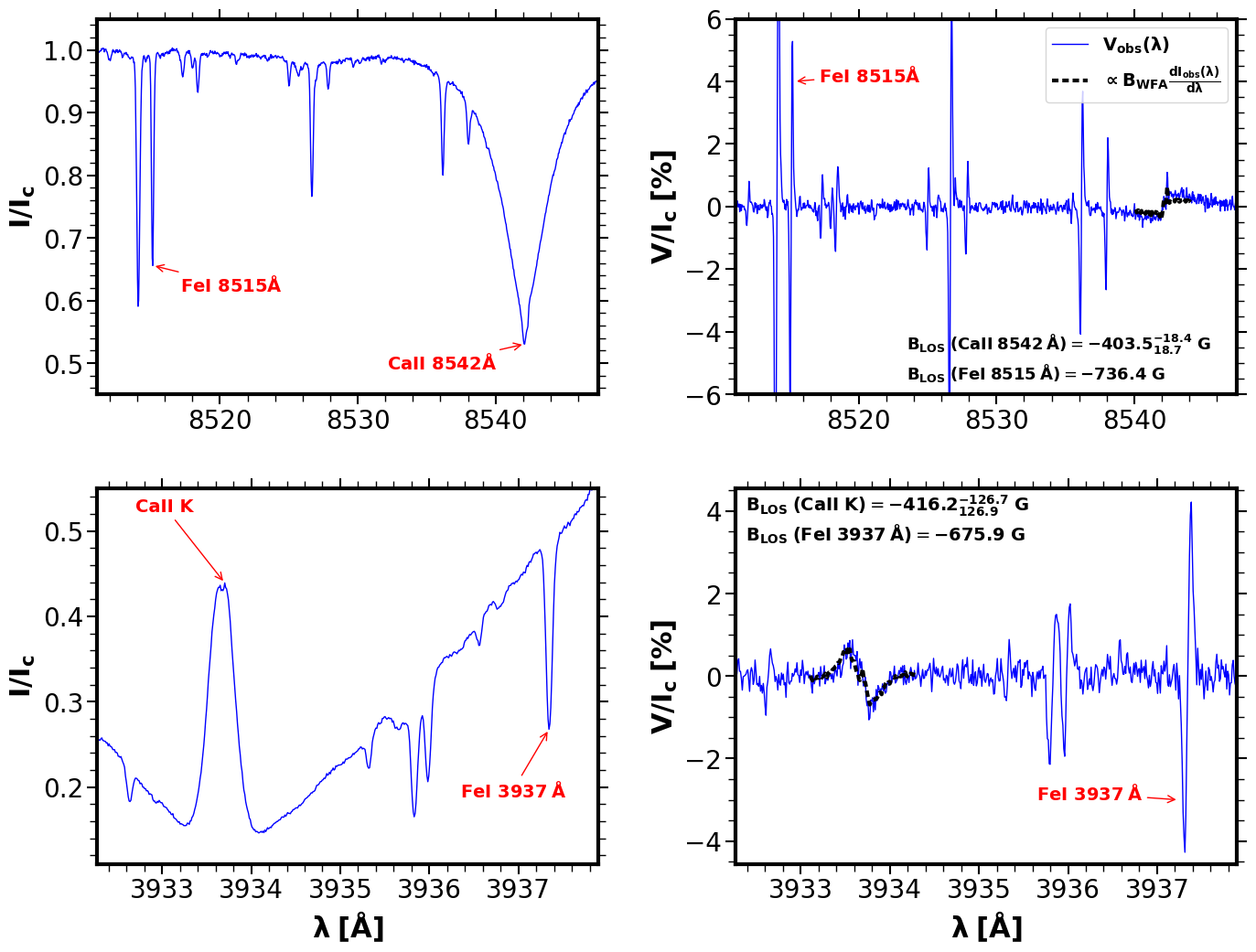}
    \caption{Stokes intensity ($I/I_{\mathrm{c}}$) and circular polarization ($V/I_{\mathrm{c}}$) profiles, normalized to the local continuum intensity, for a pixel in region A (Figures \ref{SCIP_32_PLAG_SUSI_32_PLAG_1_stokes_maps} and \ref{SCIP_32_PLAG_SUSI_32_PLAG_1_B_los_WFA_maps}) are shown in the left and right panels, respectively. Upper panels display SCIP data near the Ca~{\sc{ii}} $8542$ \AA\ line, and the lower ones show SUSI data around the Ca~{\sc{ii}} K line. In the intensity panels, we mark the four spectral lines considered in this work. In the right panels, we show the best-fit profiles (dashed black lines) in the spectral range where the WFA was applied for the Ca~{\sc{ii}} lines, also normalized to the continuum. The longitudinal components of the magnetic field inferred for the considered Ca~{\sc{ii}} and Fe~{\sc{i}} lines are also displayed in the right panels. The values from the Ca~{\sc{ii}} lines are accompanied by the upper and lower uncertainties limiting a $95\%$ confidence interval, as determined from the MCMC WFA method.}
    \label{SCIP_32_PLAG_SUSI_32_PLAG_1_point_A_I_V}
\end{figure*}

Figure~\ref{SCIP_32_PLAG_SUSI_32_PLAG_1_point_A_I_V} shows representative $I/I_{\mathrm{c}}$ and $V/I_\mathrm{c}$ profiles in a region of strong magnetic field (region A), whose position is indicated in Figures~\ref{SCIP_32_PLAG_SUSI_32_PLAG_1_stokes_maps} and~\ref{SCIP_32_PLAG_SUSI_32_PLAG_1_B_los_WFA_maps}. The SCIP and SUSI profiles are not extracted from exactly the same spatial location, as the SCIP pixel area is approximately ten times larger than that of SUSI. In this region, the Ca~{\sc{ii}} K line exhibits emission in its core, whereas the other lines remain in absorption. The circular polarization signals are strong, exceeding $5\%$ in the Fe~{\sc{i}} lines and ranging between $1\%$ and $3\%$ in the Ca~{\sc{ii}} lines. The diversity of profile shapes and amplitudes across the dataset is significant, and additional examples are provided in Appendix~\ref{Appendix_wave_int_range}.\\

Although the COG- and WFA-based inferences assume that $B_{\mathrm{LOS}}$ is constant within the formation region of the spectral line under consideration, the different $B_{\mathrm{LOS}}$ values inferred for the various considered lines highlight the vertical stratification of the magnetic field. In illustrative region A, the Fe~{\sc{i}} $8515$ \AA\ line yields $B_{\mathrm{LOS}} \approx -740 \: \mathrm{G}$, while the Fe~{\sc{i}} $3937$ \AA\ line gives $B_{\mathrm{LOS}} \approx -680 \: \mathrm{G}$. The chromospheric Ca~{\sc{ii}} $8542$ \AA\ and K lines both yield values around $-400 \: \mathrm{G}$.\\

The WFA satisfactorily reproduces the observed circular polarization profiles of the Ca~{\sc{ii}} lines. In SCIP, the WFA fit largely reproduces the observations, although small asymmetries are not fully captured. In SUSI, despite the lower S/N, the fit remains robust. The uncertainties derived from the MCMC analysis, corresponding to $95\%$ confidence intervals (Appendix~\ref{Appendix_MCMC}), are typically below $10\%$ for the Ca~{\sc{ii}} $8542$ \AA\ line, and reach up to $\sim 30\%$ for Ca~{\sc{ii}} K. The condition for applicability of the WFA, $g_{\mathrm{eff}}\frac{\Delta \lambda_{B}}{\Delta \lambda_{D}} << 1$ (see Appendix~\ref{Appendix_formation_heights}), is not strictly satisfied for the Ca~{\sc{ii}} $8542$ \AA\ at all points of the field of view, and may slightly impact the inferred $B_{\mathrm{LOS}}$. On the other hand, we do not expect it to significantly affect inferences based on the Ca~{\sc{ii}} K line.\\

As noted above, for the Fe~{\sc{i}} lines we relied on the COG technique rather than the WFA to infer $B_{\mathrm{LOS}}$. We chose this approach because the estimated validity range for the WFA  is below the magnetic field strengths present at photospheric heights in or around ARs. As discussed in Appendices~\ref{Appendix_formation_heights} and~\ref{Appendix_COG}, considerable errors could be incurred when applying the WFA in such regions.\\

Overall, the results support a scenario in which magnetic field lines expand with height, leading to a decrease in field strength and an increase in the spatial extent of magnetic structures. Close to strong photospheric concentrations, we find that $B_\mathrm{LOS}$ decreases in height whereas, outside the photospheric network, the magnetic field expansion into the chromosphere produces fields of a few hundred gauss even where the underlying photospheric field is weak or undetected (see Appendix~\ref{Appendix_wave_int_range}). This picture is consistent with the findings of previous investigations focused on the solar photosphere (e.g., \citealt{Properties_ARs}), as well as with studies of plage regions that analyze the chromospheric magnetism using the Ca~{\sc{ii}} $8542$ \AA\ line (\citealt{Morosin_2020, Pietrow_2020, da_Silva_Santos_2023}). In particular, using DKIST observations, \citet{Judge_2024_DKIST} highlighted the spatial smoothness of the chromospheric magnetic field in a plage region and discussed its implications for coronal heating. The present results show that the magnetic structures are even smoother at the greater heights probed by the Ca~{\sc{ii}} K line than at those sampled by Ca~{\sc{ii}} $8542$ \AA. This behavior was expected on the basis of theoretical considerations and radiative-MHD simulations (see, e.g., \citealt{Bjorgen_2018}). It is also in agreement with the results of investigations based on the data from the CLASP2.1 mission \citep{CLASP_2025_Ryoko}; although in the present work slightly lower chromspheric heights were probed, this was done at a substantially higher spatial resolution.\\

\section{Conclusions}
We have investigated the LOS magnetic field in an active region using co-spatial spectropolarimetric observations from the SCIP and SUSI instruments obtained during the Sunrise III flight. The combination of the Ca~{\sc{ii}} $8542$ \AA\ and Fe~{\sc{i}} $8515$ \AA\ lines from SCIP, together with the Ca~{\sc{ii}} K and Fe~{\sc{i}} $3937$ \AA\ lines from SUSI, provides diagnostics of $B_{\mathrm{LOS}}$ over a wide range of atmospheric heights, from the photosphere to the upper chromosphere. Such diagnostics were obtained by applying the COG method to the Fe~{\sc{i}} lines and the WFA to the Ca~{\sc{ii}} lines.\\

The photospheric Fe~{\sc{i}} lines reveal strong and finely structured magnetic fields, with strengths reaching up to $1000$~G. In contrast, the chromospheric Ca~{\sc{ii}} lines show systematically weaker and more spatially diffuse magnetic fields, typically of a few hundred gauss. The Ca~{\sc{ii}} $8542$ \AA\ line exhibits localized magnetic concentrations in the plage region, whereas the Ca~{\sc{ii}} K line, formed higher in the atmosphere, displays smoother and more extended structures. These results support a scenario in which the magnetic fields expand with height from the photosphere until the upper chromosphere, entailing a decrease in field strength and an increase in the spatial extent of the magnetic structures. Consistent with theoretical expectations, this work represents one of the first direct observational confirmations that the magnetic field in a plage continues to expand at heights above those probed by Ca~{\sc{ii}} $8542$ \AA\ (see also \citealt{CLASP_2024_Li, CLASP_2025_Ryoko}), while achieving an unprecedented spatial resolution.\\

The interpretation of the Ca~{\sc{ii}} diagnostics is inherently limited by their broad formation height range, which prevents isolating contributions from distinct atmospheric layers when applying the WFA. While the WFA, applied to specific lines, can yield suitable estimates for $B_{\mathrm{LOS}}$ at specific optical depths, full spectropolarimetric inversions can provide more detailed information on the variation of the magnetic field as a function of optical depth, alongside the stratification of other thermodynamic properties such as temperature or velocities. Such inversions, either applied independently to each Ca~{\sc{ii}} line or combining both, will be explored in future work using the TIC inversion module of HanleRT-TIC \citep{HanleRT_2016, HanleRT_Mg_II, HanleRT_2022, HanleRT_2024}.

\begin{acknowledgements}
We are grateful to Tanaus\'{u} del Pino Aleman for carefully reading the manuscript and suggesting several useful improvements.
We acknowledge support from the Agencia Estatal de Investigación del Ministerio de Ciencia, Innovación y Universidades (MCIU/AEI) under grant PID2022-136563NB-I00 “Polarimetric Inference of Magnetic Fields,” co-funded by the European Regional Development Fund (ERDF). Sunrise III is supported by funding from the Max-Planck-Förderstiftung (Max Planck Foundation), NASA under Grants \#80NSSC18K0934 and \#80NSSC24M0024 (“Heliophysics Low Cost Access to Space” program), and the ISAS/JAXA Small Mission-of-Opportunity program and JSPS KAKENHI Grant Numbers JP18H05234 and JP23K25916. This research has received financial support from the European Union’s Horizon 2020 research and innovation program under grant agreement No. 824135 (SOLARNET) and No. 101097844 (WINSUN) from the European Research Council (ERC). It has also been funded by the Deutsches Zentrum für Luft- und Raumfahrt e.V. (DLR, grant no. 50 OO 1608). The Spanish contributions have been funded by the Spanish MCIN/AEI under projects RTI2018-096886-B-C5, and PID2021-125325OB-C5, and from "Center of Excellence Severo Ochoa" awards to IAA-CSIC (SEV-2017-0709, CEX2021-001131-S), all co-funded by European REDEF funds, "A way of making Europe". \\
\end{acknowledgements}

\newpage

\begin{appendix}

\section{Markov chain Monte Carlo method} \label{Appendix_MCMC}

Under the WFA, the intensity and circular polarization profiles of a spectral line are related as follows \citep{Landi_2004},

\begin{equation} \label{Eq_WFA}
    V(\lambda) = -4.67 \cdot 10^{-13} g_{\mathrm{eff}} \, \lambda^{2}_{0} \, B_{\mathrm{LOS}} \frac{\partial I (\lambda)}{\partial \lambda} \; ,
\end{equation}

\noindent where $g_{\mathrm{eff}}$ and $\lambda_{0}$ are the effective Land\'e factor and the line-center wavelength of the transition (in \AA), respectively. $B_{\mathrm{LOS}}$ is the projection of the magnetic field along the LOS (in G), and $\frac{\partial I (\lambda)}{\partial \lambda}$ is the wavelength derivative of the intensity profile. The wavelength ranges around the line core considered for the application of the WFA are listed in Table~\ref{Table_wave_int_range}.\\

\begin{table}[htbp!]
    \caption{Wavelength ranges (in \AA) around the line center of each spectral line considered for the application of the WFA and the COG methods.}
    \label{Table_wave_int_range}
    \small
    \centering
    \begin{tabular}{c c c c c c}
    \hline
    & Fe~{\sc{i}} $8515$ \AA & Ca~{\sc{ii}} $8542$ \AA\ & Fe~{\sc{i}} $3937$ \AA\ & Ca~{\sc{ii}} K \\
    $\Delta \lambda$ [\AA] & $\pm 0.4$ & $\pm 2.0$ & $\pm 0.12$ & $\pm 0.6$ \\
    \hline
    \end{tabular}
\end{table}

We used an MCMC method to determine the value of $B_{\mathrm{LOS}}$ that provides the best fit between $V$ and $\frac{\partial I (\lambda)}{\partial \lambda}$ for each spectral line and at each pixel. In this context, Equation~\ref{Eq_WFA} can be written as

\begin{equation}
    V(\lambda_{i}) = a \frac{\partial I(\lambda_{i})}{\partial \lambda} + b + \epsilon_{i} \; ,
\end{equation}

\noindent where $a$ and $b$ are the slope and intercept of the linear relation, respectively, and $\epsilon_{i}$ represents Gaussian instrumental noise. For each pixel, we estimated the standard deviation of $V$, $\sigma_{V}$, from continuum wavelengths around $3946$ \AA\ in SUSI and $8505$ \AA\ in SCIP. Assuming Gaussian uncertainties with no wavelength dependence, the likelihood function is

\begin{equation}
    \mathcal{L}(a, b) = \prod_{i} \frac{1}{\sqrt{2\pi} \: \frac{\sigma_{V}}{\sqrt{2}}} \mathrm{exp} \left[ -\frac{(a\frac{\partial I_{i}}{\partial \lambda} + b - V_{i})^{2}}{2 \left( \frac{\sigma_{V}}{\sqrt{2}}\right)^{2}} \right] \; .
\end{equation}

\noindent where $i$ runs over the wavelength samples included in the fitting interval for each spectral line. The corresponding logarithmic likelihood is

\begin{equation}
    \mathrm{ln} \: \mathcal{L}(a, b) = \sum_{i} \left[-\frac{(a\frac{\partial I_{i}}{\partial \lambda} + b - V_{i})^{2}}{2\left( \frac{\sigma_{V}}{\sqrt{2}}\right)^{2}} - \frac{1}{2} \mathrm{ln}(2\pi) - \mathrm{ln} \left( \frac{\sigma_{V}}{\sqrt{2}} \right)\right] \; .
\end{equation}

We sampled the posterior distribution using the affine-invariant ensemble MCMC sampler implemented in the \texttt{emcee} package \citep{emcee}. For each fit, we used 40 walkers and 2500 steps, discarding the first 625 steps as burn-in.\\

The best-fit values of $a$ and $b$ were taken as the mean of the posterior distribution, which was found to be approximately Gaussian. The LOS magnetic field is then obtained as

\begin{equation}
B_{\mathrm{LOS}} = - \frac{a}{4.6687\cdot10^{-13} \: g_{\mathrm{eff}} \: \lambda_{0}^{2}} \; .
\end{equation}

We also estimated upper and lower uncertainties from the 2.5th and 97.5th percentiles of the posterior distribution, corresponding to a $2\sigma$ confidence interval. The intervals shown in Figure~\ref{SCIP_32_PLAG_SUSI_32_PLAG_1_point_A_I_V} therefore correspond to a $95\%$ confidence interval.\\

We verified that the $B_{\mathrm{LOS}}$ values inferred with the MCMC method are in very good agreement with those obtained from the least-squares approach described in \citet{Centeno_WFA}, with differences below $10\%$ in all pixels. Although the MCMC approach is computationally slower, we adopted it here because it provides physically meaningful uncertainty estimates.\\

\section{Formation heights of the spectral lines and range of applicability of the WFA} \label{Appendix_formation_heights}

In order to suitably interpret the height variation of the magnetic field through the multiline approach used in this work, it is necessary to understand which atmospheric layers are probed by each spectral line. We synthesized the Stokes profiles of the lines of interest with HanleRT-TIC \citep{HanleRT_2016, HanleRT_2022}, adopting a plage-like atmosphere corresponding to model P of \citet{FAL_C}, hereafter FAL-P, and a line of sight at disk center ($\mu_{\mathrm{LOS}} = 1$). The Fe~{\sc{i}} lines were treated under local thermodynamic equilibrium (LTE), whereas the Ca~{\sc{ii}} lines were computed in non-LTE. At each wavelength, we took the geometrical height at which the optical depth equals unity as the estimate for the formation height. We note that the FAL-P model is not expected to fully reproduce the properties in any given region in a solar plage. Indeed, the observed intensity profiles display considerable variation across the FOV. The resulting height estimates based on the FAL-P model should thus be regarded as indicative rather than exact.\\

\begin{figure*}[htbp!]
    \centering
    \includegraphics[width=0.65\textwidth]{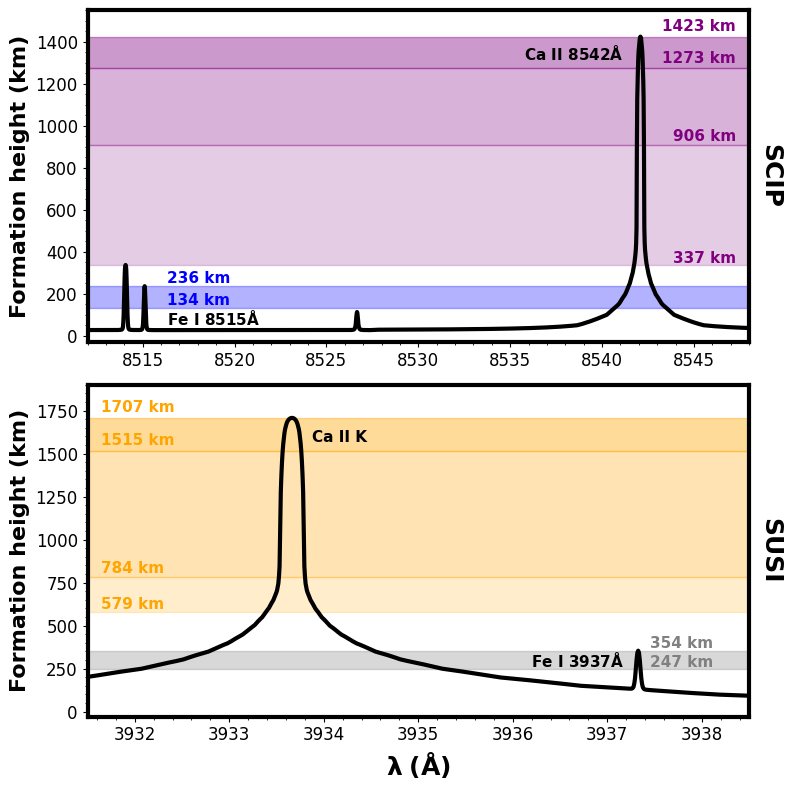}
    \caption{Estimate of the formation height as a function of wavelength, defined as the geometric height where $\tau = 1$ for a disk-center line of sight ($\mu = 1$) and computed with HanleRT-TIC for the FAL-P atmospheric model. The upper and lower panels correspond to wavelength windows comparable to those shown for SCIP and SUSI, respectively, in Figure~\ref{SCIP_32_PLAG_SUSI_32_PLAG_1_point_A_I_V}. For the Ca~{\sc{ii}} lines considered in the analysis, we highlight with colored stripes the intervals between the $\tau = 1$ heights of the following spectral features, ordered from larger to smaller height: the line core, the peaks of the inner and outer circular polarization lobes, and the continuum of the circular polarization. For the Fe~{\sc{i}} lines, only the intervals between the heights of the line core and the peaks of the $V/I_{\mathrm{c}}$ lobes are shown. All height values are indicated above their corresponding horizontal line. The Fe~{\sc{i}} $8514$ \AA\ and $8526$ \AA\ lines were also included in the calculations, at the blue and red sides of Fe~{\sc{i}} $8515$ \AA, respectively.}
    \label{Formation_heights_FALP}
\end{figure*}

The resulting formation heights are shown in Figure~\ref{Formation_heights_FALP} for the same spectral intervals as displayed in Figure~\ref{SCIP_32_PLAG_SUSI_32_PLAG_1_point_A_I_V}. For the Ca~{\sc{ii}} lines, we highlight the heights corresponding to the formation of the line core, the peaks of the inner and outer circular polarization lobes, and the continuum of the circular polarization. We define the latter as the closest wavelength to the line core where $V/I_{\mathrm{c}} < 5 \cdot 10^{-3}$ and $\frac{\partial \:(V/I_{\mathrm{c}})}{\partial \lambda} < 10^{-3}$, taking the wavelength in \AA. For the Fe~{\sc{i}} lines, only the heights corresponding to the line core and the circular polarization lobes are highlighted, as these lines display a simpler profile shape.\\

The Fe~{\sc{i}} lines used in this work probe different layers of the upper photosphere. For Fe~{\sc{i}} $8515$ \AA, the circular polarization lobes and the line core form at approximately $\sim 130$ km and $\sim 230$ km, respectively. The corresponding features in Fe~{\sc{i}} $3937$ \AA\ form about $\sim 115$ km higher. These two lines therefore provide information on distinct but relatively close photospheric layers.\\

The Ca~{\sc{ii}} lines sample a much broader range of chromospheric heights. In Ca~{\sc{ii}} $8542$ \AA, the wavelength interval between the outer circular polarization lobes encodes information from roughly $\sim 900 - 1420$ km, with the inner lobes forming around $\sim 1270$ km. In Ca~{\sc{ii}} K, the outer circular polarization lobes originate near $\sim 780$ km, whereas the inner lobes and the line core form at higher layers, around $\sim 1500 - 1700$ km, thus above the corresponding formation heights of Ca~{\sc{ii}} $8542$ \AA. This illustrates the diagnostic potential of different wavelength regions of the Ca~{\sc{ii}} K line for probing distinct chromospheric layers. A more detailed discussion is presented in Appendix~\ref{Appendix_wave_int_range}.\\

The WFA is derived under certain assumptions (e.g., \citealt{Landi_2004}), which must be  met to a reasonable degree in order for it to represent a good approximation. Firstly, it assumes that the $B_{\mathrm{LOS}}$ is constant along the formation region of each spectral line, which is more clearly satisfied for the Fe~{\sc{i}} lines than the Ca~{\sc{ii}} lines, due to their narrower formation height ranges. Although $B_{\mathrm{LOS}}$ decreases more rapidly with height in the photosphere than in the chromosphere, one may still expect a greater variation within the much broader formation range of the Ca~{\sc{ii}} lines.\\

Another necessary condition is that the Zeeman splitting produced by the magnetic field is much smaller than the Doppler width of the spectral line \citep{Landi_2004}, that is, when $g_{\mathrm{eff}}\frac{\Delta \lambda_{B}}{\Delta \lambda_{D}} \ll 1$, where

\begin{equation}\label{Eq_WFA_condition}
\Delta \lambda_{B} = 4.67 \times 10^{-13}\,\lambda_{0}^{2}\,B \label{Eq_WFA_condition_a} \quad \quad ; \quad \quad \Delta \lambda_{D} = \frac{\lambda_{0}}{c} \sqrt{\frac{2 k_{\mathrm{B}} T}{\mu M} + \xi^{2}} \; ,\label{Eq_WFA_condition_b}
\end{equation}

\noindent and $g_{\mathrm{eff}}$ and $\lambda_{0}$ are defined below Equation~\ref{Eq_WFA}, and $B$ is the absolute field strength. Here, $c$, $k_{\mathrm{B}}$, and $M$ are the speed of light, the Boltzmann constant, and the atomic mass unit, respectively. $\mu$ is the atomic weight of the atom producing the spectral line, and $T$ and $\xi$ are the temperature and microturbulent velocity of the atmosphere, respectively.\\

To estimate the magnetic field range over which the WFA is applicable, for each Ca~{\sc{ii}} line we took the values of $T$ and $\xi$ from the FAL-P model at the heights indicated by the colored horizontal lines in Figure~\ref{Formation_heights_FALP}. These correspond to the aforementioned estimates for the formation height at the line core, the inner and outer $V/I_{\mathrm{c}}$ lobes, and the $V/I_{\mathrm{c}}$ continuum. For the Fe~{\sc{i}} lines, we considered only the line core and the circular polarization lobes. For the various spectral ranges of each line — corresponding to different heights — we computed the value of $B$ for which $g_{\mathrm{eff}} \frac{\Delta\lambda_B}{\Delta\lambda_D} = 0.5$, following \citet{Afonso_CLASP_2023}. These values represent approximate upper limits for the magnetic field for which the WFA can be reasonably applied. The resulting $B$ values are listed in Table~\ref{Table_form_height_B}.\\

We note that such limits refer to full magnetic field strength and not to its longitudinal component. Thus, the limits of the applicability may often be lower in terms of $B_\mathrm{LOS}$. On the other hand, we expect that the largest $B_\mathrm{LOS}$ found in the maps presented in this work correspond to regions where the field is largely vertical (and therefore almost longitudinal). In such regions, the aforementioned upper limits should thus apply to $B_\mathrm{LOS}$ as well as $B$.\\

\begin{table}[htbp!]
    \caption{Values of $B$ for which $g_{\mathrm{eff}}\frac{\Delta \lambda_{B}}{\Delta \lambda_{D}} = 0.5$ in a plage-like (FAL-P) atmospheric model.}
    \label{Table_form_height_B}
    \small
    \centering
    \begin{tabular}{c c c c c c}
    \hline
    $B_{\mathrm{WFA, lim}}$ & Fe~{\sc{i}} $8515$ \AA\ & Ca~{\sc{ii}} $854$2 \AA\ & Fe~{\sc{i}} $3937$ \AA\ & Ca~{\sc{ii}} K \\
    \hline
    line center & 763 G & 733 G & 608 G & 1598 G \\
    inner lobes & 780 G & 720 G & 619 G & 1530 G \\
    outer lobes & & 671 G &  & 1321 G \\
    $V$ cont & & 602 G && 1241 G \\
    \hline
    \end{tabular}
\end{table}

Table~\ref{Table_form_height_B} indicates that the WFA is expected to be applicable to the Ca~{\sc{ii}} K line across the full FOV. The Ca~{\sc{ii}} $8542$ \AA\ maps show values above the estimated limits only in the strongest magnetic patches, while most of the observed region remains within the WFA regime. By contrast, the Fe~{\sc{i}} lines yield field strengths over $1000$ G, which exceeds the corresponding limits in Table~\ref{Table_form_height_B} by about $\sim 400$~G. Thus, we instead apply the COG method to infer the $B_\mathrm{LOS}$ from the Fe~{\sc{i}} lines, as discussed in Appendix~\ref{Appendix_COG}.\\

\section{The center-of-gravity method} \label{Appendix_COG}

The COG technique can be applied when the assumption of a constant magnetic field over the formation region of the considered spectral lines is approximately satisfied (see, e.g., \citealt{Landi_2004}). Unlike the WFA, it does not impose restrictions on the magnetic field strength related to the Zeeman splitting. We use it in this work to infer $B_{\mathrm{LOS}}$ from the photospheric Fe~{\sc{i}} lines.\\

\begin{figure*}[htbp!]
    \centering
    \includegraphics[width=0.9\textwidth]{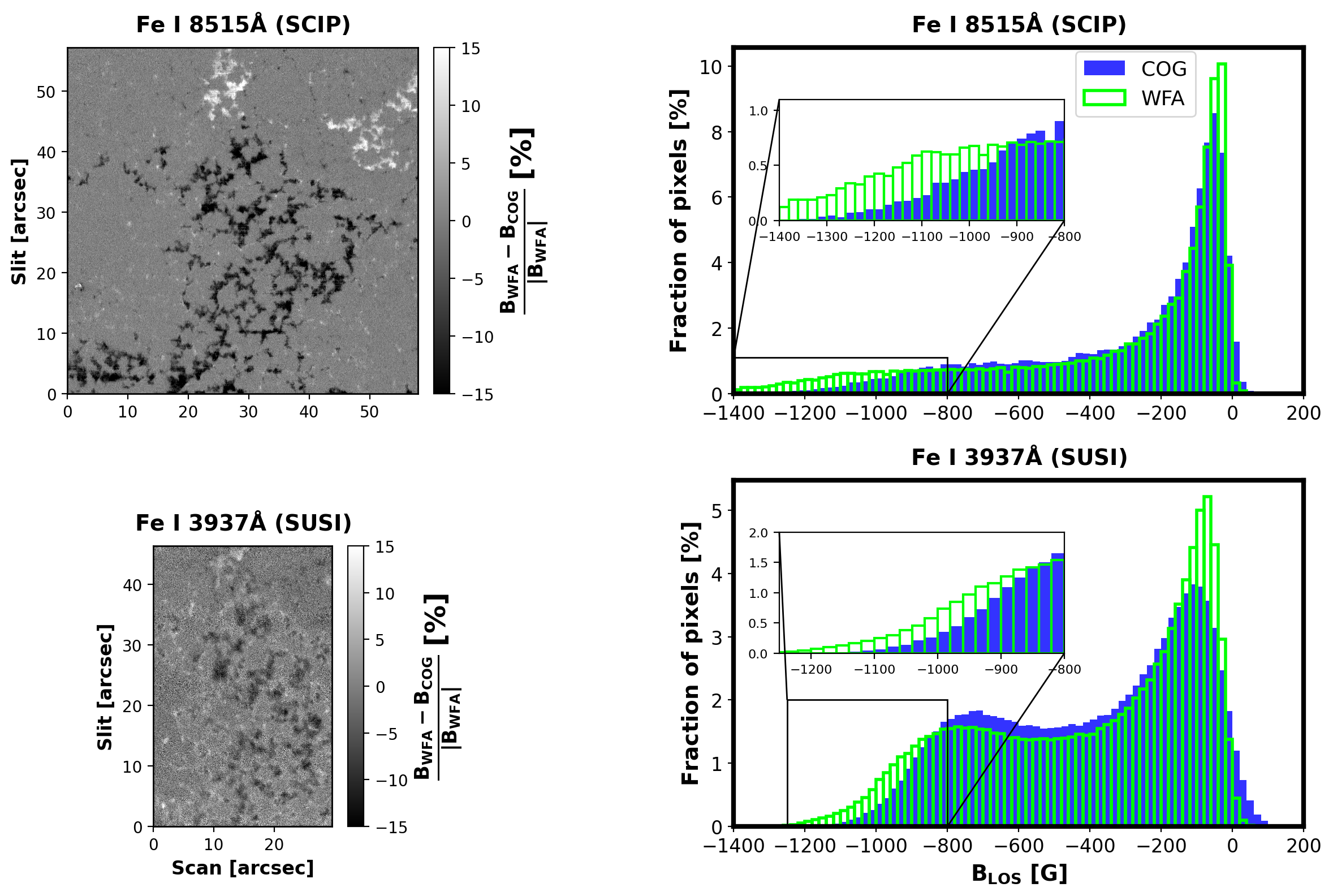 }
    \caption{Left panels: maps of the relative difference, $(B_{\mathrm{WFA}} - B_{\mathrm{COG}}) \: / \: |B_{\mathrm{WFA}}|$, expressed as a percentage. Right panels: distributions of $B_{\mathrm{WFA}}$ and $B_{\mathrm{COG}}$ within the region enclosed by the blue rectangle in Figure~\ref{SCIP_32_PLAG_SUSI_32_PLAG_1_B_los_WFA_maps}. The COG histograms are identical to those shown in the left panels of Figure~\ref{SCIP_32_PLAG_SUSI_32_PLAG_1_B_los_histograms}.}
    \label{SCIP_32_PLAG_SUSI_32_PLAG_1_B_los_COG_WFA_maps_histograms}
\end{figure*}

Within this framework,

\begin{equation}
B_{\mathrm{LOS}} = \frac{1.071 \times 10^{9}}{g_{\mathrm{eff}} \lambda^{2}_{0}} \left(\lambda^{(g)}_{+} - \lambda^{(g)}_{-}\right) \quad \quad ; \quad \quad \lambda^{(g)}_{\pm} = \frac{\int \left[\frac{1}{2} I_{\mathrm{wings}} - I_{\pm} (\lambda)\right] \lambda \: d\lambda}{\int \left[\frac{1}{2}I_{\mathrm{wings}} - I_{\pm} (\lambda)\right] \: d\lambda} \; ,
\end{equation}

\noindent with $g_{\mathrm{eff}}$ and $\lambda_{0}$ already defined below Equation~\ref{Eq_WFA}, and $I_{\pm} (\lambda) = \frac{1}{2} \left[I(\lambda) \pm V(\lambda) \right]$. We compute $I_{\mathrm{wings}}$ for each spectral line and pixel as the average of the intensity in the bluemost and redmost wavelength points within the spectral range used in the analysis (see Table~\ref{Table_wave_int_range}), where $V \approx 0$. This provides a local continuum reference consistent with the line wings in each case. However, note that it does not necessarily match the 'true' continuum intensity value, specially in SUSI, where the Fe~{\sc{i}} line is formed in the absorption wings of Ca~{\sc{ii}} K.\\

In order to assess the differences between the two techniques, we computed $B_{\mathrm{LOS}}$ using both of them, namely $B_{\mathrm{WFA}}$ and $B_{\mathrm{COG}}$. The left panels of Figure~\ref{SCIP_32_PLAG_SUSI_32_PLAG_1_B_los_COG_WFA_maps_histograms} show maps of the relative difference between the inferred field strengths. These closely resemble the spatial patterns seen in the Fe~{\sc{i}} maps of Figure~\ref{SCIP_32_PLAG_SUSI_32_PLAG_1_B_los_WFA_maps}. The differences reach up to $\sim 15\%$ in the strongest field patches of the Fe~{\sc{i}} $8515$ \AA\ map, while generally remaining below $\sim 10\%$ in Fe~{\sc{i}} $3937$ \AA. Although one could expect the discrepancies to be larger for the case of the Fe~{\sc{i}} $3937$ \AA\ line, because of its lower threshold for applicability of the WFA (see Table \ref{Table_form_height_B}), we also note that this line forms somewhat higher in the photosphere, where the magnetic fields are generally weaker (as shown in Fig~\ref{SCIP_32_PLAG_SUSI_32_PLAG_1_B_los_WFA_maps}).\\

The right panels of Figure~\ref{SCIP_32_PLAG_SUSI_32_PLAG_1_B_los_COG_WFA_maps_histograms} show the distributions of $B_{\mathrm{WFA}}$ and $B_{\mathrm{COG}}$ within the unipolar region enclosed by the blue rectangles in Figure~\ref{SCIP_32_PLAG_SUSI_32_PLAG_1_B_los_WFA_maps}. In the strongest field regions, the WFA systematically yields larger values than the COG method, as shown by the more extended tails of the $B_{\mathrm{WFA}}$ distributions toward larger $|B_{\mathrm{LOS}}|$ values.\\

\section{Wavelength range for the WFA in the Ca~{\sc{ii}} K line} \label{Appendix_wave_int_range}

\begin{figure*}[htbp!]
    \centering
    \includegraphics[width=0.85\textwidth]{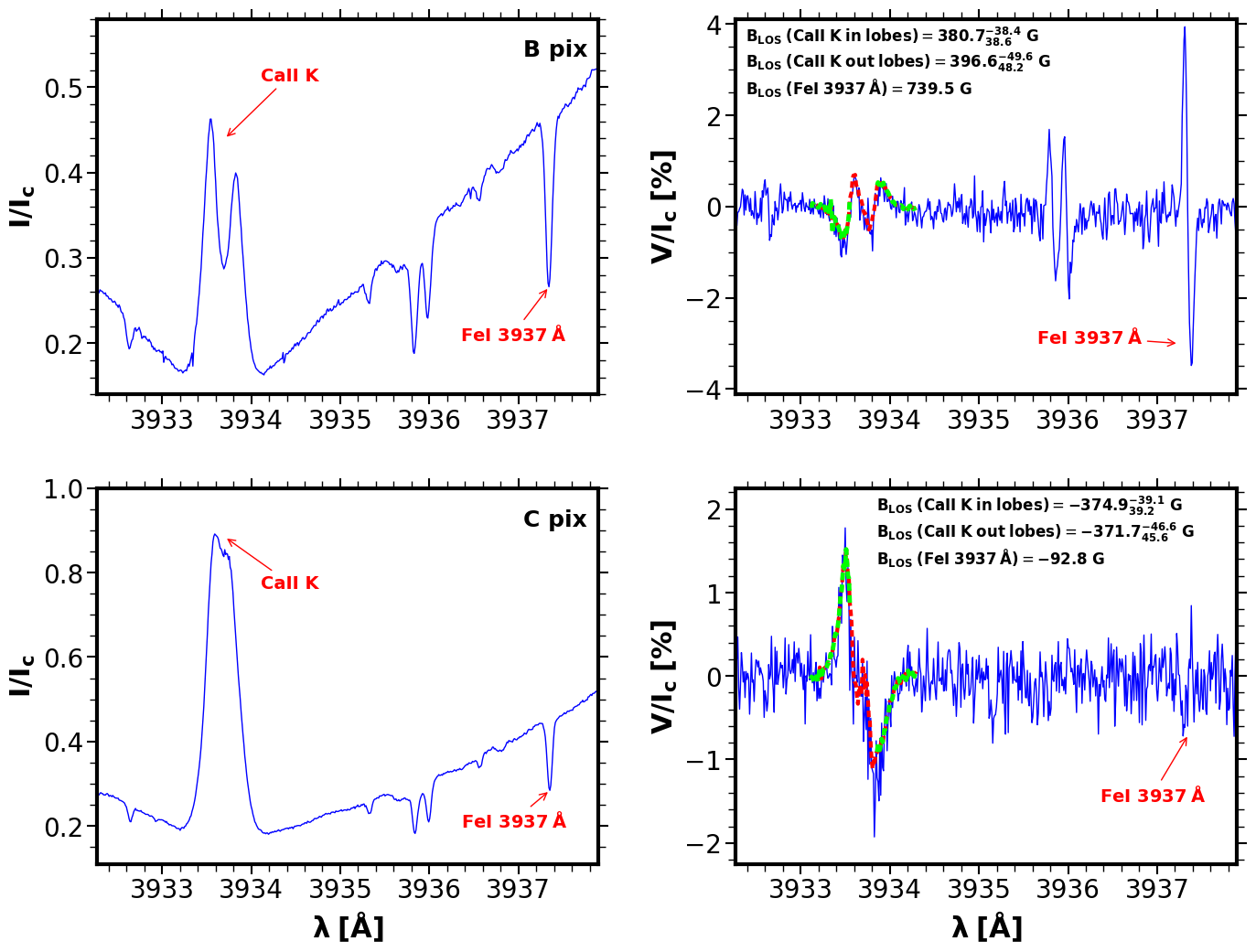 }
    \caption{Stokes intensity ($I/I_{\mathrm{c}}$) and circular polarization ($V/I_{\mathrm{c}}$) profiles, normalized to the local continuum intensity, are shown in the left and right panels, respectively. Upper and lower panels display the profiles corresponding to regions B and C, respectively, marked with orange dots in Figures~\ref{SCIP_32_PLAG_SUSI_32_PLAG_1_stokes_maps} and \ref{SCIP_32_PLAG_SUSI_32_PLAG_1_B_los_WFA_maps}. In the intensity panels, we mark the four spectral lines considered in this work. In the right panels, we plot the best-fit profiles in the inner ($\Delta \lambda = \pm 0.15$ \AA, dashed red lines) and outer ($\Delta \lambda = \pm [0.15, 0.6]$ \AA, dashed green lines) lobes where the WFA was applied to the Ca~{\sc{ii}} lines. The longitudinal components of the magnetic field inferred for the considered Ca~{\sc{ii}} and Fe~{\sc{i}} lines are also displayed in the right panels. The values from the Ca~{\sc{ii}} lines are accompanied by the upper and lower uncertainties limiting a $95\%$ confidence interval, as determined from the MCMC WFA method.}
    \label{SUSI_32_PLAG_1_points_B_C_I_V}
\end{figure*}

As discussed in Appendix~\ref{Appendix_formation_heights}, selecting different wavelength intervals within a given spectral line allows one to probe different atmospheric layers. This approach has already been applied to other chromospheric diagnostics, for example to the Mg~{\sc{ii}} h\&k lines observed in the CLASP missions \citep{CLASP_2021_Ryoko, Review_TB_dPA, Afonso_CLASP_2025}.\\

For the Ca~{\sc{ii}} K line, applying the WFA only to the inner circular-polarization lobes makes it possible to infer information more representative of the upper chromosphere, reducing the contribution from lower layers. By contrast, the outer lobes encode information from lower heights. We note, however, that applying the WFA to the outer lobes of the K line can lead to underestimated magnetic field values \citep{IJB_2025}, in a way similar to that found for the Mg~{\sc{ii}} doublet \citep{HanleRT_2022}. In Figure~\ref{SUSI_32_PLAG_1_points_B_C_I_V}, we show the intensity and $V/I_{\mathrm{c}}$ profiles for two locations in the SUSI dataset: region B, a positive-polarity strong-field pixel inside the photospheric network, and region C, a negative-polarity pixel outside the photospheric network. Both locations are marked with orange dots in Figures~\ref{SCIP_32_PLAG_SUSI_32_PLAG_1_stokes_maps} and \ref{SCIP_32_PLAG_SUSI_32_PLAG_1_B_los_WFA_maps}.\\

The Ca~{\sc{ii}} K intensity profile in region B exhibits a clear self-reversal at line center. In the circular polarization profile, the inner lobes are clearly distinguishable and have amplitudes comparable to those of the outer lobes. We therefore applied the WFA in two separate spectral intervals in order to infer magnetic fields representative of different heights: a $\Delta \lambda = \pm 0.15$ \AA\ range around the line core, containing mainly the inner lobes, and a $\Delta \lambda = \pm [0.15, 0.6]$ \AA\ range containing only the outer lobes. The inferred values are $B_{\mathrm{LOS}} \sim 380$ G and $\sim 400$ G for the inner and outer ranges, respectively, consistent with a decrease of magnetic field strength with height in the chromosphere. However, the inferred uncertainties are much larger than said difference. In both intervals, the WFA provides a satisfactory fit to the observed $V/I_{\mathrm{c}}$ profile.\\

In region C, the Ca~{\sc{ii}} K emission profile is slightly asymmetric and does not show an evident self-reversal. The inner circular-polarization lobes are present, but are much less pronounced than in region B. This is expected under the WFA, since the circular polarization is proportional to the wavelength derivative of the intensity profile. The inferred $B_{\mathrm{LOS}}$ values are very similar in the inner and outer spectral ranges, namely $\sim -375$ G and $\sim -370$ G, respectively. In this region, the LOS magnetic field inferred from the Fe~{\sc{i}} line is close to zero because the pixel lies outside the photospheric network. This is fully consistent with the spatial distribution discussed in Section~\ref{Section_results}.\\

However, applying the WFA only to the inner circular polarization lobes is not feasible over the full FOV, because in many pixels their amplitudes are too small to ensure a sufficiently high S/N ratio. This is evident in the profiles corresponding to regions A and C (lower panels in Figures~\ref{SCIP_32_PLAG_SUSI_32_PLAG_1_point_A_I_V} and \ref{SUSI_32_PLAG_1_points_B_C_I_V}). As a consequence, maps derived exclusively from the inner lobes do not show the degree of smoothness expected for upper-chromospheric magnetic fields. For this reason, in the present work we applied the WFA to the full spectral range containing the inner and outer lobes together with the line core. Spatial and/or spectral binning could enhance the signal in the inner $V/I_{\mathrm{c}}$ lobes and enable a more height-selective analysis, but this lies beyond the scope of the present study.\\
\end{appendix}

\bibliography{bibliography}{}
\bibliographystyle{aasjournalv7}

\end{document}